\documentclass[twocolumn,aps,prl,amssymb,amsmath,superscriptaddress,floatfix,showpacs]{revtex4}
\usepackage{graphicx}
\usepackage{epstopdf}
\usepackage{times}

\def\TC{\textit{T}$_{\mathrm{c}}$}
\def\R7byR3{$\sqrt{7}\times\sqrt{3}$}

\begin{document}

\title{Radial Band Structure of Electrons in Liquid Metals}

\author{Keun Su Kim}\email[Present address: Advanced Light Source, Lawrence Berkeley National Laboratory, Berkeley, CA 94720, USA.]{}\affiliation{Center for Atomic Wires and Layers, Pohang University of Science and Technology, Pohang 790--784, Korea}

\author{Han Woong Yeom}\email[yeom@postech.ac.kr]{}\affiliation{Center for Atomic Wires and Layers, Pohang University of Science and Technology, Pohang 790--784, Korea}\affiliation{Department of Physics, Pohang University of Science and Technology, Pohang 790--784, Korea}

\date{\today}

\begin{abstract}

The electronic band structure of a liquid metal was investigated by measuring precisely the evolution of angle-resolved photoelectron spectra during the melting of a Pb monolayer on a Si(111) surface. We found that the liquid monolayer exhibits a free-electron-like band and it undergoes a coherent radial scattering, imposed by the radial correlation of constituent atoms, to form a characteristic secondary hole band. This unique \textit{double radial bands} and their gradual evolution during melting can be quantitatively reproduced, including detailed spectral intensity profiles, with our radial scattering model based on a theoretical prediction of 1962. Our result establishes the radial band structure as a key concept for describing the nature of electrons in strongly disordered states of matter.

\end{abstract}


\maketitle

%--- Introduction1 ---
Central to understanding various electronic properties of crystalline materials is the band theory based on periodic electron waves. In reality, however, a huge variety of materials are subject to significant disorders to disturb periodic crystalline structures \cite{LEE,MOTT,NELSON,STRAND}, making it challenging to describe their electronic structures \cite{PETROFF,BAUMBERGER,VOIT,ROTENBERG}. Most notoriously, the nature of electrons in a liquid metal (LM) without any long-range order \cite{STRAND,NELSON,PETROFF,BAUMBERGER} has been extremely difficult to understand and no well-established idea exists so far to connect their atomic and electronic structures.

%--- Introduction2 ---
In recent years, important progresses were made to understand the atomic structure of LMs, establishing the local configurations of liquid fragments \cite{REICHERT,CHEN,NEGU,KAMINSKI} and their global radial correlation \cite{REICHERT,CHEN,NEGU,KAMINSKI,GREY,REED,HENZLER,LIU}. In sharp contrast, while various theoretical ideas were proposed since 1960's to describe the electronic structure of LMs \cite{EDWARDS,BALLEN,ZIMAN,STOLL,HALDER}, critical experimental information to develop these ideas has been lacking. To our knowledge, there  is only one experimental report to directly map out the band structure of LMs with angle-resolved photoemission (ARPES) \cite{BAUMBERGER}. It used a liquid Pb monolayer supported on a crystalline Cu(111) surface, where the momentum (\textit{k}) resolved information can be preserved in the photoemission process \cite{BAUMBERGER}. This work found two persistent Fermi surfaces (FSs) to have very short electron coherence lengths in LMs. However, the origin of these two bands and their markedly distinct coherence lengths have been mysterious. In the experimental point of view, this limitation was partly due to (\textit{i}) a too high melting temperature ({\TC} = 568 K) to resolve the fine FS shape and (\textit{ii}) the intermixed signal from the metallic substrate.

%--- Our Work ---
In this Letter, we overcome such limitations by choosing a Pb monolayer on a semiconducting Si(111) substrate with a band gap, which has anomalously low {\TC} $\sim$ 270 K \cite{HWANG,TRINGIDES1,CHOI} owing to its intrinsic diffusion property \cite{ALTMAN,TRINGIDES2}. This low {\TC} allows us to trace the fine evolution of ARPES spectra in the melting process at a relatively low temperature with reduced thermal broadening. We identify the electronic fingerprint of coherent radial scattering, which is a missing link to relate the atomic and electronic structures in LMs. Our radial scattering model based on an old theoretic prediction \cite{EDWARDS} explains the origin of unique \textit{double radial bands} as well as their mysterious coherence lengths, providing comprehensive understanding of the electronic band structure of LMs.

%--- Figure 1: Structure & LEED ---
\begin{figure}[b]
\includegraphics[width=8 cm]{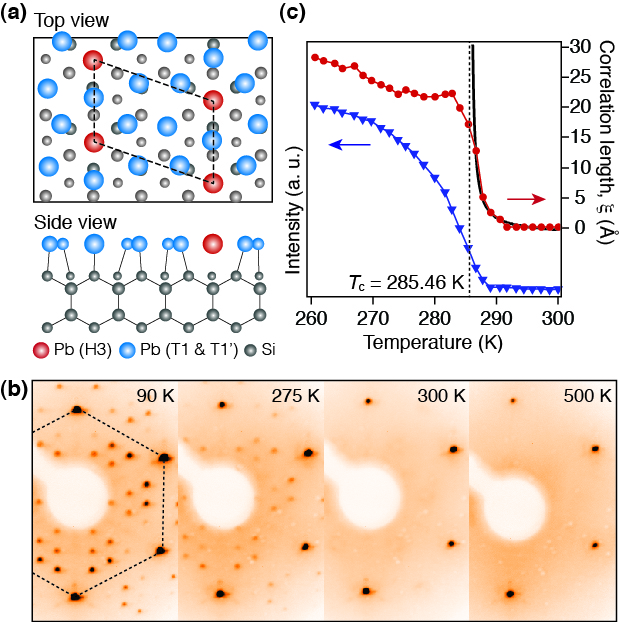}
\caption{(color online). (a) Crystalline structure of the Pb monolayer on Si(111) with the areal density of 9.4 Pb atoms nm$^{-2}$ \cite{JUNG}. Dashed lines represent the {\R7byR3} surface unit cell. (b) LEED patterns taken at marked temperatures with an electron energy of 81 eV. Dashed lines indicate Si(111)1$\times$1 spots. (c) Intensity and correlation length as a function of temperature around {\TC} obtained by the peak-fitting analysis of a {\R7byR3} superspot in (b) using the correlation function $\propto$ exp(--$|$\textit{r}$|$/$\xi$), \textit{r} $\rightarrow$ $\infty$ \cite{HENZLER}. The solid line is a fit with the critical exponent of 0.37 predicted in a 2D liquid theory \cite{NELSON}.}\label{Fig1}
\end{figure}

%--- Experiments ---
The \textit{n}-type Si(111) wafer was thermally cleaned in the ultra-high vacuum chamber (6.5 $\times$ 10$^{-11}$ torr). The Pb monolayer was fabricated by \textit{in-situ} thermal deposition of Pb onto the Si(111)7$\times$7 surface held at 500 K. The melting transition is completely reversible and does not depend on the annealing history up to 700 K. The density of metal deposits was thoroughly calibrated with the well-known phase diagram \cite{TRINGIDES1}. ARPES spectra were taken using a Scienta R4000 electron analyzer and a high-flux He discharge lamp (He I${\alpha}$, 21.2 eV). The energy and angular resolutions were 10 meV and 0.25$^{\circ}$, respectively.

%--- Structure & Melting ---
Figure 1(a) shows the crystalline structure below {\TC}. The Pb layer has two groups of atoms, one sitting on top of surface Si atoms (blue, T1 or T1$^{\prime}$) and the other at the hollow sites in between (red, H3) \cite{TRINGIDES1,CHOI}. These result in the {\R7byR3} periodicity [dashed lines in Fig. 1(a)] with respect to Si(111)1$\times$1. The melting of the Pb monolayer is demonstrated in the temperature series of diffraction patterns [Fig. 1(b)]. In clear contrast with the Si-related 1$\times$1 spots (dashed lines), the higher order superspots from the Pb monolayer diminish rapidly at around room temperature. The quantitatively measured intensity of Pb-related spots in Fig. 1(c) reveals more details of the temperature dependence. The correlation length ($\xi$) extracted from the diffraction width \cite{HENZLER} shows an abrupt downturn to reach to a few {\AA}. This value is on the order of the interatomic spacing, which manifests the liquid state. The curve fitting of $\xi$(\textit{T}) with the critical exponent value of 0.37 predicted in the 2D melting theory \cite{NELSON} follows well the experimental results [solid line in Fig. 1(c)], and determines the {\TC} of 285.5 K.

%--- Figure 2: Spectral evolution ---
\begin{figure}
\includegraphics[width=8 cm]{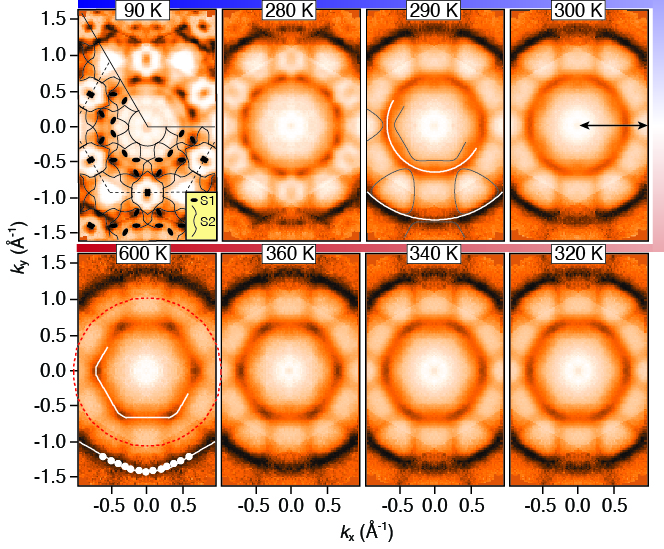}
\caption{(color online). Experimental FS data at various marked temperatures. Overlaid in 90 K data is the band theory calculaton \cite{KIM,JUNG} and dashed lines indicate the Si(111)1$\times$1 Brillouin zone. Across {\TC} $\sim$ 285 K, the complex FS reduces into a large outer and a smaller inner FSs (white lines), the latter of which gradually transforms from circle to hexagon as further increasing temperature. The other weak features (grey lines) are due to remaining substrate potentials. The white dots at 600 K are band-peak positions obtained by fitting the radial spectral profiles. The red dots show the midpoints of inner and outer FSs in the radial direction.}\label{Fig2}
\end{figure}

%--- FS1 ---
Figure 2 shows the temperature series of ARPES intensity maps at the Fermi energy. The intensity map taken at 90 K, far below {\TC}, represents the FS of the solid Pb monolayer \cite{KIM}, which can be well reproduced by the band theory calculation (overlaid lines) \cite{JUNG}. It consists of two metallic bands S1 and S2 woven complexly according to the {\R7byR3} periodicity within the silicon band gap, which originate from in-plane Pb 6\textit{p} orbitals \cite{JUNG,KIM}.

%--- FS2 ---
As approaching {\TC} at 280 K, the complex FS pattern gradually disappears and is dominated at 290 K by two simple circular FSs, a large outer and a small inner one (white lines). Very similar FSs were also observed on Pb/Cu(111) \cite{BAUMBERGER}, suggesting that this double-band structure is characteristic to the liquid Pb monolayer. Upon further heating to 300--360 K, these FSs undergo reversible shape changes, gradually from circles to rounded hexagons, which is more obvious for the inner one. This finding, which was overlooked in the previous study \cite{BAUMBERGER}, gives an important insight into the mechanism behind the double bands as discussed below. There are other, much weaker, features of triangles and hexagons (grey lines), which are due to the weak 1$\times$1 substrate potential [Fig. 1(b)] \cite{SUP}. These extra features decay gradually into the background at higher temperatures, whereas the double hexagonal FSs, associated with the liquid monolayer, become more dominant in 600 K data, indicating their distinct origins. The observed double-hexagon topology of FSs, along with its apparent shape change with temperature, is not compatible at all with any periodic band model of a crystalline solid, calling for a new description beyond the scope of Bloch waves.

%--- Figure 3: Model calculation ---
\begin{figure}
\includegraphics[width=8 cm]{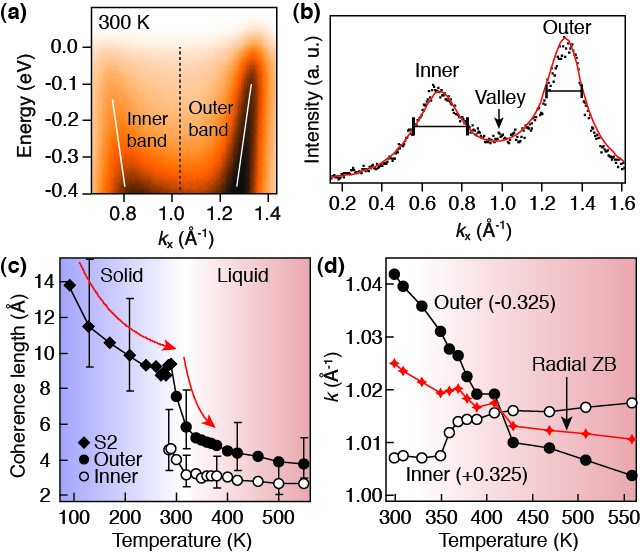}
\caption{(color online). (a) Band dispersions taken along the arrow in Fig. 2. (b) Typical \textit{k} distribution curve taken from (a) at the Fermi energy. The red line overlaid is the result of the model calculation in Fig. 4. Temperature dependence of (c) electron coherence length and (d) Fermi wavevector of the outer (closed circles) and the inner (open circles) bands. The coherence length is obtained from the spectral \textit{k} width. The Fermi wavevectors of inner and outer bands are offset respectively by $\pm$0.325 {\AA}$^{-1}$. Red dots in (d) show the center of inner and outer Fermi wavevectors (radial zone boundary).}\label{Fig3}
\end{figure}

%--- Band ---
Figure 3(a) shows the band dispersions taken along the arrow in Fig. 2, which indicate that the outer and the inner FSs are electron and hole pockets, respectively. The outer band with a dominant intensity disperses following a free-electron-like parabola, and can be traced to nearly free Pb-valence electrons with limited coherence \cite{BAUMBERGER,VOIT}, which is a reasonable approximation for electronic states away from the band gap in the theoretical model of LMs \cite{BALLEN,ZIMAN,HALDER}. Supporting this idea, the measured electron density of the outer band (\textit{n} = 2.98 $\times$ 10$^{15}$ cm$^{-2}$) amounts to the whole free-electron density of the Pb monolayer. This, however, leaves the origin of the inner band mysterious. One important finding is that the outer and the inner bands disperse symmetrically; these two bands can exactly be overlapped by folding at the middle \textit{k} point [dashed line in Fig. 3(a)]. This denies the previous assumption that these two bands originate from different orbitals \cite{BAUMBERGER}. The band folding is typical for electron waves modulated by a given periodic or aperiodic structural order \cite{VOIT,ROTENBERG}, and suggests some structural order of LMs.

%--- Theoretical idea ---
Motivated by a theoretical idea \cite{EDWARDS}, we focus on the remaining structural order in LMs, the radial atomic correlation, which has been well-established with diffractions \cite{REICHERT,GREY,REED,HENZLER}. This early theory considers simple plane waves scattered by screened ionic potentials. The range of ionic order (\textit{R}) versus electron mean-free path ($\lambda$) determines the qualitative nature of electron scattering. There are two extreme cases: When \textit{R} is much larger than $\lambda$, the situation is similar to polycrystals or glassy metals \cite{EDWARDS}, in which many tiny crystalline domains are distributed in random fashion. The electrons would be globally incoherent yet locally close to Bloch states. In this case, the primary band would be translated according to the local crystalline periodicity but the lost orientational order distributes these translation vectors uniformly in all directions as in Fig. 4(a). These uniform translations of the primary FS would overlap to yield the secondary FS-like feature or a broad hole pocket. The other extreme case is an ideal liquid, where \textit{R} is comparable (or shorter) to $\lambda$ \cite{EDWARDS}. In this case, the electrons are subject to a radial potential due to the radial atomic correlation, leading to scattering preferentially along the radial direction. Thus, the primary band would be radially scattered to create a more clear hole band as in Fig. 4(b), which constitutes the characteristic double-band structure. Although the distinction between these two cases in reality may not be too obvious, the fundamental difference is the electron scattering nature, uniform or radial scattering. For a LM at a finite temperature, the electron state would be in between these two extremes. Moreover, as \textit{R} tends to decrease with increasing temperature, the band evolution above {\TC} would be a crossover in between these two extremes with the gradually enhanced radial scattering.

 %--- Figure 4: Graphs ---
\begin{figure}
\includegraphics[width=8 cm]{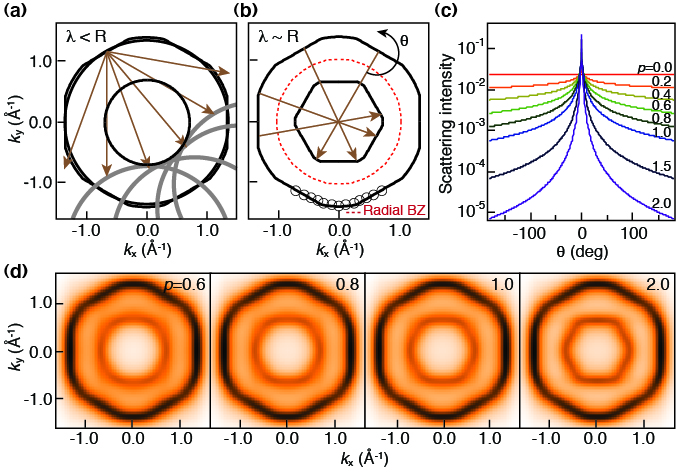}
\caption{(color online). Schematic illustrations of the electron scattering and expected FSs for two models in Ref. \cite{EDWARDS}, (a) the polycrystalline model and (b) the ideal liquid model. The arrows represent examples of possible scattering vectors and the open circles are experimental band positions at 600 K. (c) Angular distribution of scattering intensity (in log scale) from the radial direction used for (d). The degree of radial scattering is parameterized by the \textit{p} value. (d) Calculated FSs based on the simple model considering the gradually enhanced radial scattering [that is, varying the \textit{p} value in (c)].}\label{Fig4}
\end{figure}

%--- FS modelling1---
Based on the above idea, we performed a simple model calculation \cite{SUP}. In our model, the enhancement of the radial scattering is parameterized by the power (\textit{p}) value that determines the angular decay of scattering intensity from the radial direction [Fig. 4(c)]; uniform scattering at \textit{p} = 0 and strong radial scattering at higher \textit{p} values. We determined the primary band shape from the curve-fitting analysis of the outer FS at 600 K (white dots in Fig. 2). This model successfully reproduces the major experimental observations [Fig. 4(d)]; the characteristic double-band structure, and more importantly the gradual shape change of the inner FS from circle at \textit{p} = 0.6 to hexagon at \textit{p} = 2.0. This remarkable agreement between the model and the experiment corroborates the validity of our interpretation and the old theory \cite{EDWARDS}; the hole band comes from the radial scattering of the primary band.

%, analogous to the periodic folding of Bloch bands in a crystalline solid

%--- FS modelling2 ---
The gradual shape change of the inner FS is a consequence of enhanced radial scattering, indicating the continuous electronic evolution above {\TC} from incoherent Bloch waves to radial waves. This can be related to continuous melting that passes through an intermediate glassy state \cite{REED,LIU}. The hexagonal shape of the primary FS at a higher temperature, which is not very clear at the intermediate state due to overlapped 1$\times$1 bands (Fig. 2), was also consistently found in liquid Pb/Cu(111) \cite{BAUMBERGER}. This reflects a local point symmetry of liquid Pb fragments, another important structural characteristics of LMs \cite{REICHERT,CHEN,KAMINSKI,NEGU}. While the hexagonal symmetry itself can be intrinsic to the liquid Pb monolayer \cite{REICHERT,KAMINSKI}, their orientational alignment would be preferentially along the underlying substrate potential mesh \cite{NEGU,GREY}. This allows the observation of the apparent band structure from LMs \cite{BAUMBERGER,REICHERT}. Even under the finite substrate influence, the electron wave only coherent over a few atoms [Fig. 3(c)] represents a strongly disordered electronic system, and would be indistinguishable from an ideal 2D liquid \cite{BAUMBERGER}.

 %--- Substrate effect --- 
The electron coherence length as a function of temperature is displayed in Fig. 3(c). Above {\TC} the inner band has a significantly, about 1.5 times, shorter coherence length (that is, broader spectral width) than the outer one. This is more clearly manifested in the \textit{k} distribution curve in Fig. 3(b). A consistent observation was made for the liquid Pb/Cu(111), where the authors interpreted it as a different degree of localization and argued the distinct orbital symmetry and/or the wavelength of two bands as possible origins \cite{BAUMBERGER}. However, according to our interpretation the inner band is just a pronounced trace of complex scatterings from the outer primary band, and its broader spectral width is natural and intrinsic to the scattered origin. Indeed, our model calculation does reproduce the distinct spectral widths and even the strange valley intensity in between [the red line in Fig. 3(b) and see \cite{SUP} for the \textit{y} direction]. In addition, the same procedure can be successfully extended to all binding energies throughout the experimental dispersions \cite{SUP}. That is, our radial scattering model provides a description for the whole spectral function of LMs.

%--- Radial BZ ---
For the ideal liquid model in Fig. 4(b), one can introduce the concept of \textit{radial Brillouin zone} \cite{EDWARDS,STOLL}, as counterparts of periodic Brillouin zone in crystalline solids \cite{VOIT} and `quasi-Brillouin zone' in quasicrystals \cite{ROTENBERG}. Connecting the midpoints of observed inner and outer bands in every radial direction, one can experimentally trace the distribution of the reference \textit{k} points for the radial scattering. The measured midpoints are surprisingly isotropic (radial) as shown by red dots in Fig. 2. This is in clear contrast to the conventional Brillouin zone of a simple polygon for a 2D solid \cite{PETROFF}. We term the double-radial band and the radial Brillouin zone as \textit{the radial band structure}, which is the key concept for the electronic structure of LMs. This would have wide implications not only in developing more sophisticated theory but in understanding exotic electronic properties of various strongly disordered systems, for example, electron localization in glassy metals \cite{MOTT, LEE}.

%--- Quantitative info ---
From the \textit{k} position of the radial Brillouin zone ($|$\textit{q}$|$/2), one can derive the radial scattering amplitude ($|$\textit{q}$|$) and the average interatomic spacing (\textit{a} = 2$\pi$/$|$\textit{q}$|$) \cite{CHEN}. In Fig. 3(d), we show the temperature dependence of radial zone boundary. It decreases with increasing temperature and converges to about 1.01 {\AA}$^{-1}$, which yields $|$\textit{q}$|$ = 2.03 {\AA}$^{-1}$ and \textit{a} = 3.10 {\AA}. In fact, this is fully consistent with the known values for liquid Pb/Ge(111) in X-ray diffraction (\textit{a} = 3.06 {\AA}) \cite{GREY} and our own analysis for Pb/Cu(111) (\textit{a} = 3.11 {\AA}) \cite{BAUMBERGER}. The remarkable agreement on three independent systems with different surface lattice constants (3.61--4.00 \AA) points to the existence of universal interatomic spacing for liquid Pb fragments irrespective of substrates, indicating dominant interaction within a 2D liquid. The radial Brillouin zone is thus a powerful concept to link atomic and electronic structures of LMs.

%--- Acknowledgement ---
This work was supported by NRF through the CRi program. We thank M. H. Kang and H. J. Choi for helpful discussions.

\end{document}